\documentclass[a4paper,preprintnumbers,amsmath,amssymb,nofootinbib,10pt]{revtex4}

\usepackage[margin=2cm]{geometry}
\usepackage{amssymb}
\usepackage{mathrsfs}
 \usepackage{graphicx}

\begin{document}

\title{Cosmological constraints on a decomposed Chaplygin gas}
\author{Yuting Wang$^{1,}$$^{2}$, David Wands$^{2}$, Lixin Xu$^{1}$, Josue De-Santiago$^{3,}$$^{2}$ and Alireza Hojjati$^4$}

\affiliation{$^1$School of Physics and Optoelectronic Technology,
Dalian University of Technology, Dalian, Liaoning 116024, People's
Republic of China \\
$^2$Institute of Cosmology and Gravitation, University of
Portsmouth, Portsmouth, PO1 3FX, United Kingdom \\
$^3$ Universidad Nacional Aut\'onoma de M\'exico, 04510, Distrito Federal,
M\'exico \\
$^4$ Institute for the Early Universe, Ewha Womans University,
Seoul, 120-750, South Korea}

\date{\today}
\begin{abstract}
Any unified dark matter cosmology can be decomposed into dark matter
interacting with vacuum energy, without introducing any additional
degrees of freedom. We present observational constraints on an
interacting vacuum plus dark energy corresponding to a generalised
Chaplygin gas cosmology. We consider two distinct models for the
interaction leading to either a barotropic equation of state or dark
matter that follows geodesics, corresponding to a rest-frame sound
speed equal to the adiabatic sound speed or zero sound speed,
respectively. For the barotropic model, the most stringent
constraint on $\alpha$ comes from the combination of CMB+SNIa+LSS(m)
gives $\alpha<5.66\times10^{-6}$ at the 95\% confidence level, which
indicates that the barotropic model must be extremely close to the
$\Lambda$CDM cosmology. For the case where the dark matter follows
geodesics, perturbations have zero sound speed, and CMB+SNIa+gISW
then gives the much weaker constraint $-0.15<\alpha<0.26$ at the
95\% confidence level.
\end{abstract}
\maketitle

\section{Introduction}

The dominance of dark energy and dark matter in the present cosmological energy density has led many authors to seek a unified description in terms of a single dark component, referred to as unified dark matter or quartessence \cite{quartessence}, which should explain both the current accelerated expansion of the Universe and the role of nonbaryonic dark matter in structure formation.
One candidate from this kind of model, which has attracted a lot of
attention, is the generalized Chaplygin gas (GCG) model
\cite{gcg1,gcg2}. It is a fluid, defined by an exotic equation of
state,
\begin{eqnarray}
 \label{PgCg}
P_{gCg}=-\frac{A}{\rho_{gCg}^\alpha} \,,
\end{eqnarray}
with model parameters $\alpha$ and $A$. The GCG model with
$\alpha=1$ reduces to the original Chaplygin gas model
\cite{gcg1,cg}.
Predictions of the GCG cosmology has been tested against the different
observational data, including type Ia supernovae (SNIa)
\cite{sngcg1,sngcg2,sngcg3}, the positions of cosmic microwave
background (CMB) radiation peaks \cite{cmbgcg1,cmbgcg2,cmbgcg3}, the X-ray
gas mass fraction of clusters \cite{Xraygcg1}, Hubble
parameter-redshift data \cite{hgcg1,hgcg2}, the lookback
time-redshift data \cite{lookgcg1}, gamma-ray bursts \cite{gmbgcg1},
and as well as various combinations of the updated data
\cite{jointgcg1}.

Going beyond constraints provided by the homogeneous background
cosmology, the inclusion of inhomogeneous linear perturbations in
the GCG has an effect on both the cosmic microwave background
anisotropies and structure formation. In particular, the imprint of
a large integrated Sachs-Wolfe (ISW) signal \cite{isw}, which arises
from a time-dependent gravitational potential when dark energy
dominates at late times in the Universe, could be observed on the
CMB anisotropies \cite{Carturan:2003, Amendola:2003}. Recently, in Ref.~\cite{Xu:2012} two of us revisited constraints on the GCG model from
the combination of CMB, baryon acoustic oscillations (BAO), and SNIa, including the full CMB temperature and
polarization power spectra data. A tight bound for the parameter
$\alpha$ was obtained up to the order of $10^{-3}$.

The effect of the GCG on structure formation also contains important
information. In Ref.~\cite{Sandvik:2002jz}, it was argued that in order
to predict large-scale structure (LSS) consistent with the data from
the 2dF survey \cite{2dF}, the GCG model has to be indistinguishable
from $\Lambda$CDM, with $\alpha<\mathcal{O}(10^{-5})$. The slightly
looser constraint, $\alpha<\mathcal{O}(10^{-4})$ was obtained in Ref.~\cite{Park:2009np} by fitting the predicted baryon matter power
spectrum, rather than the total matter power spectrum, to the Sloan
Digital Sky Survey (SDSS) DR7 power spectrum data\footnote{The
theoretical baryon power spectrum may also be compatible with SDSS
observations for large $\alpha>1$, corresponding to a superluminal sound speed
\cite{Gorini:2008}, however the CMB power spectrum does not favour this regime \cite{Park:2009np}.}.
In either case a stringent limit is placed on the $\alpha$ parameter because there are either instabilities or oscillations
in the GCG power spectra due to the adiabatic pressure perturbation
produced for nonzero $\alpha$.
%
This problem can be avoided in some unified dark matter models by
requiring a fast transition between a CDM-like era and a
$\Lambda$CDM-like phase \cite{Piattella:2009kt, Bertacca:2010mt,
Bruni:2012}.

In order to save the GCG model some type of intrinsic entropy perturbation must be
introduced in the GCG fluid to counteract the effect caused by the adiabatic GCG pressure
perturbation,
as in the so-called silent GCG model \cite{inNAgcg}. The combined constraints from data on
both the silent GCG and standard GCG model are presented in Ref.~\cite{Luca:2005}. It is shown that with a vanishing GCG pressure
perturbation there is a much wider allowed parameter range, $-0.3<\alpha<0.7$
\cite{Luca:2005}. Subsequently, the ISW data \cite{iswdata} was used
to constrain
the silent GCG model \cite{gcgisw}.
The required entropy perturbation can be introduced by decomposing
the GCG fluid into two components of dark matter and dark energy
\cite{NAgcg}, but the possible ways of decomposition are not unique
\cite{Chimento:2010}. In Ref.~\cite{Bento:2004uh} the GCG model was split
into dark matter and vacuum energy, considering a homogeneous vacuum
when studying density perturbations in the Newtonian limit.

Recently it has been shown that any unified dark matter model can be decomposed into dark matter interacting with a vacuum energy (with equation of state $P=-\rho$) without introducing any additional degrees of freedom \cite{Wands:2012}.
In general an interacting vacuum energy is inhomogenous, and a description of the dynamics of inhomogeneous vacuum energy was presented in Refs.~\cite{Wands:2012,inprep}.

In this paper we will study in detail the constraints on a decomposed GCG model described by dark matter interacting with inhomogeneous vacuum energy. A closed set of perturbation equations are obtained once a covariant form is specified for the interaction.
Firstly, we consider a model in which the vacuum energy is a function of the local matter density. We show this leads to a barotropic equation of state for the dark matter+vacuum, and hence reproduces previous results for a unified GCG. Secondly we consider an energy-momentum transfer parallel to the matter four-velocity, leading to a geodesic flow for the matter.
In both cases we consider background cosmologies that mimic the GCG cosmologies, but the evolution of linear perturbations differs due to the presence of non-adiabatic pressure perturbations.
We pay particular attention to their imprints on the CMB and LSS power spectra. To constrain the barotropic and geodesic model parameters we will use
different combinations of current cosmological data including the
CMB temperature and polarization power spectra \cite{wmapdata7},
SNIa from the Union2.1 compilation of the Supernova Cosmology
Project Collaboration \cite{ref:sn580}, the BAO distance
measurements \cite{ref:Percival2}, the LSS power spectrum from SDSS
DR7 \cite{ref:LSS}, and the galaxy-ISW (gISW) cross-correlation
power spectra from Ref.~\cite{ref:gISW}.

This paper is organized as follows. In the next section, we review
the background equations for the GCG model. In Sec.~III, the linear
perturbation equations for both the unified GCG model and the two
decomposed models are presented. In Sec.~IV, we show the effects on
CMB and LSS power spectra in the two decomposed models, then perform
a global fitting to observational data by using the Markov Chain
Monte Carlo (MCMC) method and discuss the resulting constraints on the interaction. In the final
section we present our conclusions. In what follows (unless
otherwise specified) the illustrative plots are shown using the
joint mean values for cosmological parameters from WMAP7 data
\cite{wmapdata7}: $\Omega_bh^2=0.02255, \Omega_{dm}h^2=0.1126,
h=0.702, \tau=0.088, n_s=0.968, \textmd{and} ~A_s=2.43\times10^{-9}~
\textmd{at}~ k_{s0}=0.002 \textmd{Mpc}^{-1}$.


\section{Chaplygin gas cosmology}

We will consider a spatially flat, homogeneous and isotropic Friedmann-Robertson-Walker
universe with metric
\begin{eqnarray}
ds^2=-dt^2+a^2(t)\delta_{ij}dx^idx^j \,,
\end{eqnarray}
where the Friedmann equation is given by
\begin{eqnarray}
 \label{Friedmann}
H^2\equiv\left(\frac{\dot{a}}{a}\right)^2=\frac{8 \pi
G}{3}(\rho_b+\rho_r+\rho_{gCg}) \,,
\end{eqnarray}
where $\rho_b$, $\rho_r$ and $\rho_{gCg}$ are the energy densities of
baryons, radiation and a generalised Chaplygin gas which can act as both
a dark energy component, accelerating the Universe at late times, and dark matter.
The energy conservation equations for the various components read
\begin{eqnarray}
\dot{\rho}_b+3H\rho_b=0 \,, \\
\dot{\rho}_r+4H\rho_r=0 \,, \\
 \label{dotrhogCg}
\dot{\rho}_{gCg}+3H(\rho_{gCg}+P_{gCg})=0 \,.
\end{eqnarray}
The equation of state for GCG (\ref{PgCg}) then allows us to
integrate Eq.~(\ref{dotrhogCg}) to obtain
\begin{eqnarray}
 \label{rhogCg}
 \rho_{gCg} = \rho_{gCg0} \left[ B_s + (1-B_s)a^{-3(1+\alpha)}
 \right]^{1/(1+\alpha)} \,,
\end{eqnarray}
where $\rho_{gCg0}$ is the present value of energy density of GCG (when
$a=a_0=1$) and we have replaced $A$ in Eq.~(\ref{PgCg}) with the
dimensionless model parameter, $B_s=A/\rho_{gCg0}^{1+\alpha}$.

In a previous work we have shown that a unified dark matter fluid
with an arbitrary equation of state can be decomposed into
pressureless dark matter interacting with a vacuum energy, $V$, with
an appropriate energy transfer \cite{Wands:2012} (see also Ref.~\cite{Chimento:2010}).
For the GCG solution (\ref{rhogCg}), we can write
\begin{eqnarray}
 \label{decompose}
\rho_{gCg} &=& \rho_{dm}+V \,,
\end{eqnarray}
such that, from Eq.~(\ref{PgCg})
\begin{eqnarray}
 \label{A}
  A&=&V(\rho_{dm}+V)^{\alpha} \,.
\end{eqnarray}
The conservation equations of dark matter and vacuum energy are
\begin{eqnarray}
 \label{rhodmdot}
\dot{\rho}_{dm}+3H\rho_{dm} &=& -Q \,, \\
 \label{Vdot}
\dot{V} &=& Q \,.
\end{eqnarray}
Thus from Eqs.~(\ref{A})-(\ref{Vdot}), we
obtain
\begin{eqnarray}
 \label{backQ}
 Q = 3\alpha H\frac{\rho_{dm} V}{\rho_{dm} +V} \,.
\end{eqnarray}

In the decomposed model (\ref{decompose}), the parameter, $B_s$ in
Eq.~(\ref{rhogCg}), can be written in terms of the present values of
the energy densities, $B_s=[V/(\rho_{dm}+V)]_0$. The decomposed
model is characterised by a single dimensionless parameter $\alpha$
\cite{Wands:2012}, but an FRW solution is specified by two boundary
conditions (such as the present density of matter and vacuum)
whereas the original GCG model is specified by two model parameters,
$\alpha$ and $A$ (or $B_s$) and one boundary condition. This already
suggests that the decomposed model may naturally accommodate
nonadiabatic density perturbations, in addition to the adiabatic
perturbations of the original GCG. As we shall see, this can have
important consequences for the evolution of perturbations in a
decomposed model.

Note that the Friedmann equation (\ref{Friedmann}) can be written as
\begin{eqnarray}
 \label{finalFriedmann}
H^2=H_0^2\left\{\Omega_{b}a^{-3}+\Omega_{r}a^{-4}+(\Omega_V+\Omega_{dm})\left[\frac{\Omega_V}{\Omega_V+\Omega_{dm}}+\frac{\Omega_{dm}}{\Omega_V+\Omega_{dm}}a^{-3(1+\alpha)}\right]^{\frac{1}{1+\alpha}}\right\}
\,,
\end{eqnarray}
where the dimensionless fractional energy density of individual
components is defined as
\begin{eqnarray}
\Omega_{I}=\left(\frac{8 \pi G \rho_{I}}{3H^2}\right)_0 \,,
\end{eqnarray}
which satisfies $\Omega_b+\Omega_r+\Omega_V+\Omega_{dm}=1$.

\begin{figure}[!htbp]
\includegraphics[width=8cm]{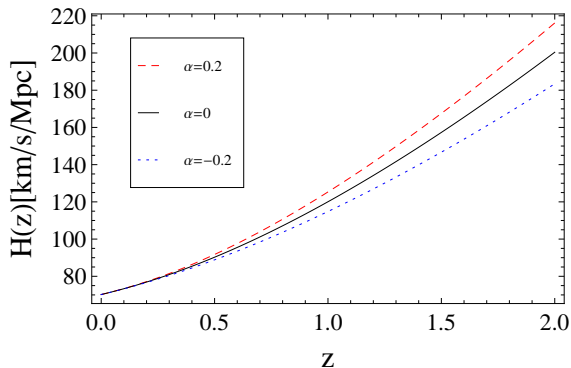}\\
  \caption{The Hubble parameter as a function of redshift z for different values of parameter $\alpha$,
  where the solid line with $\alpha$ = 0 corresponds to $\Lambda$CDM model.
 }\label{fig:H}
\end{figure}

We illustrate the background evolution as a function of redshift for
different values of the parameter $\alpha$ in Fig.~\ref{fig:H}. In
the particular case when $\alpha=0$, the background evolution is
identical to that for the $\Lambda$CDM model.

The FRW background solution, Eq.~(\ref{finalFriedmann}), shows a degeneracy between the decomposed vacuum+dark matter with the interaction (\ref{backQ}) and the unified GCG with equation of state (\ref{PgCg}), where $\Omega_{gCg}=\Omega_V+\Omega_{dm}$ and $B_s=\Omega_V/(\Omega_V+\Omega_{dm})$. As discussed in Ref.~\cite{Chimento:2010},
this is an example of a more general degeneracy between interacting and unified models. One can either specify an exotic equation of state for a unified dark matter model, or one can obtain many equivalent decomposed models containing interacting fluids with a specified energy transfer. Although equivalent to a unified model at the background level, we expect different physical models for the interaction to be distinguished by the evolution of inhomogeneous perturbations.


\section{The Linear perturbation equations}

We begin with a general scalar mode of the metric perturbation in an
arbitrary gauge with the perturbed line of element
\cite{Mukhanov:review,Malik:thesis,Malik:Wands:review,Malik:Wands:interfield}
\begin{eqnarray}
 ds^2 = -(1+2\phi)dt^2+2a\partial _i B dt dx^i
  + a^2 \left[(1-2\psi)\delta_{ij}+2\partial_i\partial_j E \right] dx^i dx^j \,.
\end{eqnarray}
The Einstein equations are
\cite{Mukhanov:review,Malik:thesis,Malik:Wands:review,Malik:Wands:interfield}
\begin{eqnarray}
 \label{time-time}
3H\left( \dot\psi +H\phi \right) - \frac{\nabla^2}{a^2} ( \psi + H
\sigma ) &=& -4\pi G \delta \rho \,,\\
 \label{time-space}
\dot\psi +H\phi &=& - 4\pi G (\rho+P) \theta \,,\\
  \label{traceless}
 \dot{\sigma}+H\sigma-\phi+\psi&=&8 \pi G \Pi \,,\\
   \label{trace}
 \ddot{\psi}+3H\dot{\psi}+H\dot{\phi}+(3H^2+2\dot{H})\phi&=& 4\pi
 G\left(\delta P+\frac{2}{3}\frac{\nabla^2}{a^2}\Pi\right) \,,
\end{eqnarray}
where $\sigma\equiv a^2\dot{E}-aB$ is the shear, $\delta\rho$ is the
total density perturbation, $\delta P$ is the total pressure
perturbation, $\Pi$ is the total anisotropic stress, and $\theta$ is
the total covariant velocity perturbation\footnote{Note that
$\theta\equiv a(v+B)$ in terms of the total velocity potential $v$
defined in, e.g., Ref.~\cite{Malik:Wands:interfield}.}.

For each individual component, the covariant conservation equation
allowing for energy-momentum transfer gives
\begin{eqnarray}
 \label{covcon}
\nabla_\mu T^{\mu\nu}_{(I)}=Q^\nu_{(I)} \,.
\end{eqnarray}
General relativity requires conservation of the total
energy-momentum, $\nabla_\mu T^{\mu\nu}=0$, so we have $\sum_I
Q^\nu_{(I)}=0$. The perturbed energy-momentum transfer four-vector
is conventionally split into the energy transfer, $Q_I+\delta Q_I$,
and momentum transfer, $f_I$, relative to the total four-velocity
\cite{Malik:Wands:review,Malik:Wands:interfield,Kodama:Sasaki:review,Malik:interfluid}
\begin{eqnarray}
 \label{QmuI}
 Q_{\mu}^{(I)} = \left[-Q_I(1+\phi)-\delta Q_I \,, \partial_i (f_I + Q_I\theta) \right]
 \,.
\end{eqnarray}
The energy and momentum conservation equations for each fluid
component are then \cite{Malik:thesis,
Malik:Wands:review,Malik:Wands:interfield, Malik:interfluid}:
\begin{eqnarray}
 \label{gendeltarho}
 \dot{\delta\rho}_I + 3H(\delta\rho_I+\delta P_I)
 -3 (\rho_I+P_I)\dot\psi + (\rho_I+P_I) \frac{\nabla^2}{a^2}( \theta_I + \sigma)
 &=& \delta Q_I +Q_I\phi \,,\\
 \label{gentheta}
(\rho_I+P_I)\dot\theta_I - 3c_{sI}^2H(\rho_I+P_I)\theta_I +
(\rho_I+P_I)\phi +\delta P_I+\frac{2}{3}\frac{\nabla^2}{a^2}\Pi_I
&=& f_I+Q_I \theta -(1+c_{s I}^2) Q_I \theta_I \,,
\end{eqnarray}
where $c_{sI}^2=\dot{ P}_I/\dot{\rho}_I$ is the adiabatic sound
speed.

The numerical code CAMB \cite{ref:cambweb} is written in the synchronous
gauge, so in order to place quantitative constraints on the model parameters we will solve the
equations in a synchronous gauge \cite{ref:Ma1995} where
\begin{eqnarray}
 \label{sg}
\phi=B=0 \,,~~~ \psi=\eta \,,~~~ \textmd{and} ~~k^2
E=-\frac{h}{2}-3\eta \,.
\end{eqnarray}

There is a residual spatial gauge freedom in the synchronous gauge
which allows one to set the initial velocity of any component to
zero. We see from Eq.~(\ref{gentheta}) that it is possible to work
in a gauge in which $\theta_I=0$ at all times for non-interacting,
pressureless matter. A comoving synchronous gauge is commonly used,
in which the dark matter velocity is zero, $\theta_{dm}=0$. This is
no longer possible in models where the dark matter experiences a
momentum transfer, $f_{dm}+Q_{dm}(\theta-\theta_{dm})\neq0$. We can
still work in a synchronous gauge, but note that the dark matter
velocity is not necessarily zero.

The time-time and time-space components of the perturbed Einstein field
equations, (\ref{time-time}) and (\ref{time-space}), in the
synchronous gauge are
\begin{eqnarray}
 \label{00sg}
\frac{\dot{h}}{2} &=& \frac{1}{H}\left(\frac{k^2}{a^2}\eta+4 \pi G
\delta\rho\right) \,, \\
 \label{0jsg}
 \dot{\eta} &=& -4 \pi G (\rho+P)\theta \,.
\end{eqnarray}
Together with the energy and momentum conservation equations for each component, these form a closed set of equations.

After decoupling, the baryon component is conserved independently and the perturbation
equations for the baryon density contrast and velocity are given by
\begin{eqnarray}
 \label{benergy-sg}
 \dot{\delta}_b -\frac{k^2}{a^2}\theta_b &=& -\frac{\dot{h}}{2} \,, \\
  \label{bmom-sg}
 \dot\theta_b &=& 0 \,.
\end{eqnarray}
Combining Eqs.~(\ref{traceless}), (\ref{trace}) in the synchronous
gauge and Eq.~(\ref{00sg}), we have
\begin{eqnarray}
\label{htwodot}
 \ddot{h}+2H\dot{h} = -8 \pi G (\delta\rho+3\delta P) \,.
\end{eqnarray}
By differentiating Eq.~(\ref{benergy-sg}) with respect to time and
combining Eq.~(\ref{bmom-sg}) and the above equation, we find a
second-order differential equation for the baryonic density contrast,
\begin{eqnarray}
\label{baryontwo}
 \ddot{\delta}_b+2H\dot{\delta}_b = 4 \pi G (\delta\rho+3\delta P) \,.
\end{eqnarray}

\subsection{Unified GCG model}

First let us consider perturbations in the unified GCG model, where the energy-momentum tensor of the GCG is conserved
and hence Eqs.~(\ref{gendeltarho}) and (\ref{gentheta}) give the perturbed energy and momentum conservation equations
\begin{eqnarray}
 \label{deltagcg}
 \dot{\delta}_{gCg} + 3H(c_{s,gCg}^2-w_{gCg}){\delta}_{gCg}
 +(1+w_{gCg})\left(-\frac{k^2}{a^2}\theta_{gCg}+\frac{\dot{h}}{2}\right)
 &=& 0 \,,\\
 \label{thetagcg}
\dot\theta_{gCg} - 3c_{s,gCg}^2H\theta_{gCg}
+\frac{c_{s,gCg}^2}{1+w_{gCg}}\delta_{gCg} &=& 0 \,,
\end{eqnarray}
where the adiabatic sound speed for the GCG is fixed by the equation
of state of the GCG
\begin{eqnarray}
\label{csgcg}
c_{s,gCg}^2 &=& -\alpha w_{gCg} \,, \\
w_{gCg} &=& -\frac{B_s}{B_s+(1-B_s)a^{-3(1+\alpha)}} \,.
\end{eqnarray}

We can derive a second-order differential equation for GCG
overdensity by differentiating Eq.~(\ref{deltagcg}) with respect to
time and combining Eqs.~(\ref{htwodot}) and (\ref{thetagcg}),
\begin{eqnarray}
 \label{gcgtwodot}
\ddot{\delta}_{gCg}&+&2H[1+3(c_{s,gCg}^2-w_{gCg})]\dot{\delta}_{gCg}
 \nonumber \\
&+&[3\dot{H}(c_{s,gCg}^2-w_{gCg})+3H(c_{s,gCg}^2-w_{gCg})^.+9H^2(c_{s,gCg}^2-w_{gCg})^2+6H^2(c_{s,gCg}^2-w_{gCg})]\delta_{gCg}
 \nonumber \\
&=&
-c_{s,gCg}^2\frac{k^2}{a^2}\delta_{gCg}
 +(1+w_{gCg})3c_{s,gCg}^2H\frac{k^2}{a^2}\theta_{gCg}
 +(1+w_{gCg})4 \pi G (\delta\rho+3\delta P) \,.
\end{eqnarray}
The first term on the right-hand-side of the above equation
dominates in the small-scale limit and the adiabatic pressure
perturbation causes the instabilities (or oscillations) of density
perturbations for negative (or positive) $\alpha$ \cite{Bean}.

\subsection{Interacting vacuum energy + dark matter models}

Let us now consider a general interacting vacuum energy + dark matter model,
where the energy continuity equations (\ref{gendeltarho}) for the dark
matter and vacuum energy in the synchronous gauge become
\begin{eqnarray}
 \dot{\delta\rho}_{dm} + 3H\delta\rho_{dm}  + \rho_{dm}\left(-\frac{k^2}{a^2} \theta_{dm} + \frac{\dot{h}}{2}\right)
 &=& \delta Q_{dm} \,,\\
 \label{deltaQV}
 \dot{\delta V} &=& \delta Q_V \,,
\end{eqnarray}
while the momentum conservation equations (\ref{gentheta}) are given by
\begin{eqnarray}
 \label{fdm}
 \rho_{dm}\dot\theta_{dm} &=& f_{dm} +Q_{dm}(\theta-\theta_{dm}) \,,\\
 \label{fV}
 -\delta V &=&  f_V+Q_V\theta \,.
\end{eqnarray}
Note that the vacuum energy has vanishing momentum and therefore the
force exerted on the vacuum, $\partial_i(f_V+Q_V\theta)$, is exactly
canceled by the gradient of the vacuum pressure,
$\partial_i(-\delta V)$.

The conservation of total energy-momentum for the interacting dark matter and vacuum
energy implies that
\begin{eqnarray}
 Q_V=- Q_{dm}=Q \,,~~~\delta Q_V=-\delta Q_{dm} \,,~~~f_V=-f_{dm} \,.
\end{eqnarray}
We can then eliminate $\delta Q_{dm}$ and $f_{dm}$ using the energy and momentum conservation equations for the vacuum
(\ref{deltaQV}) and (\ref{fV}) to leave the
energy-momentum conservation equations for the dark matter
\begin{eqnarray}
  \label{finadelta}
 \dot{\delta}_{dm} -\frac{Q}{\rho_{dm}}\delta_{dm}  -\frac{k^2}{a^2}\theta_{dm} +\frac{\dot{h}}{2}
 &=&-\frac{\dot{\delta V}}{\rho_{dm}} \, \\
  \label{finatheta}
 \rho_{dm}\dot\theta_{dm}  &=& \delta V+Q\theta_{dm} \,,
\end{eqnarray}
where the background energy transfer $Q$ is given by
Eq.~(\ref{backQ}).


Note that the interacting vacuum+dark matter introduces no additional degrees of freedom with respect to the original GCG. We have two coupled first-order equations (\ref{finadelta}) and (\ref{finatheta}) for the dark matter density and velocity, instead of the two coupled first-order equations (\ref{deltagcg}) and (\ref{thetagcg}) for the GCG density and velocity. The background solution determines the adiabatic sound speed and equation of state for the GCG, whereas a physical model for the interaction is required to determine the vacuum perturbation, $\delta V$, and can affect the sound speed for the interacting vacuum+dark matter.

Generally we can identify two classes of interacting models which could be used to specify the energy-momentum transfer in the decomposed model. Either one can specify the vacuum energy as a function, $V(X)$, of some scalar quantity $X$, and hence $\delta V=V'(X)\delta X$. Or else one can identify a four-vector, $U^\mu$, which the energy-transfer (\ref{covcon}) is parallel to: $Q_V^\mu\propto U^\mu$. In the following we will consider one example of each class of interacting model, but we note that these are not unique and it may be interesting to consider other possibilities with different observational signatures.


\subsubsection{Barotropic model}

Firstly, we consider a model in which the vacuum energy density is a function of local dark matter density
\begin{equation}
\label{vrhoc}
 V=V(\rho_{dm}) \,.
 \end{equation}
This implies that the energy transfer between the vacuum and the dark matter is along the gradient of the local dark matter density:
$Q_V^\mu\propto \nabla^\mu\rho_{dm}$.

\begin{figure}[!htbp]
\includegraphics[width=8cm]{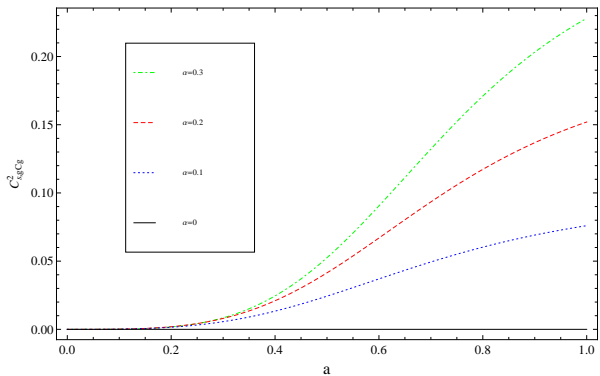}\\
  \caption{The sound speed as a function of scale factor a for the barotropic model.
 }\label{fig:cs}
\end{figure}

At linear order, we have
\begin{eqnarray}
 \label{deltav}
 \delta V=\frac{\dot{V}}{\dot{\rho}_{dm}}\delta \rho_{dm} \,,
\end{eqnarray}
which shows that in this case the relative perturbations between the
vacuum and the dark matter are adiabatic. The total pressure
perturbation for the interacting dark matter+vacuum energy is
\begin{eqnarray}
 -\delta V &=& \frac{\alpha V}{\rho_{dm}+V}(\delta\rho_{dm}+\delta
 V) \nonumber \\
 &=& c_{s,gCg}^2 (\delta\rho_{dm}+\delta
 V) \,.
\end{eqnarray}
Therefore, the sound speed for the interacting dark matter+vacuum
energy is exactly the same as the adiabatic sound speed for the
original generalised Chaplygin gas, $c_{s,gCg}^2$ given by
Eq.~(\ref{csgcg}) \cite{Wands:2012, inprep}. We show the effect on
the sound speed caused by the parameter $\alpha$ in
Fig.~\ref{fig:cs}.

In the synchronous gauge, we have from Eqs.~(\ref{finadelta}),
(\ref{finatheta}) and (\ref{deltav})
\begin{eqnarray}
\label{cenergy-sg}
\dot{\delta}_{dm} &=& \frac{(1+\alpha)Q}{\rho_{dm}+(1+\alpha)V}\delta_{dm}+\frac{\rho_{dm}+(1+\alpha)V}{\rho_{dm}+V}\frac{k^2}{a^2}\theta_{dm}-\frac{\rho_{dm}+(1+\alpha)V}{\rho_{dm}+V}\frac{\dot{h}}{2} \,, \\
\label{cmom-sg}
 \dot\theta_{dm} &=& \frac{Q}{\rho_{dm}}\theta_{dm}-\frac{\alpha V}{\rho_{dm}+(1+\alpha)V}\delta_{dm} \,,
 \end{eqnarray}
It is clear that the barotropic model with $\alpha = 0$ reduces to
the $\Lambda$CDM model at the linear order. Comparing with dark
matter perturbations in the standard $\Lambda$CDM model, we can see
that density contrasts and velocities of the interacting dark matter
evolve very differently. In this case, due to the presence of
momentum transfer, $\theta_{dm}$ will evolve to be nonzero, even if
its initial value is set to zero. Therefore, the synchronous gauge
is not comoving with the interacting dark matter and we calculate
the perturbed equations in an arbitrary synchronous gauge
\cite{arbitrarySG}.

Combining Eqs.~(\ref{cenergy-sg}) and (\ref{cmom-sg}) with
Eq.~(\ref{htwodot}), we obtain a second-order differential equation
for the dark matter overdensity
\begin{eqnarray}
\label{barcdmtwo} \ddot{\delta}_{dm} &-&
\left[\frac{2(1+\alpha)Q}{\rho_{dm}+(1+\alpha)V}-2H\right]\dot{\delta}_{dm}  \nonumber \\
&-& \left\{\left[\frac{(1+\alpha)Q}{\rho_{dm}+(1+\alpha)V}\right]^.-\left[\frac{(1+\alpha)Q}{\rho_{dm}+(1+\alpha)V}\right]^2+2H\left[\frac{(1+\alpha)Q}{\rho_{dm}+(1+\alpha)V}\right]\right\}\delta_{dm} \nonumber \\
&=& -c_{s,gCg}^2\frac{k^2}{a^2}\delta_{dm}+\left(1+\frac{\alpha
V}{\rho_{dm}+V}\right)\frac{Q}{\rho_{dm}}\frac{k^2}{a^2}\theta_{dm}+\left(1+\frac{\alpha
V}{\rho_{dm}+V}\right)4 \pi G (\delta\rho+3\delta P) \,.
 \end{eqnarray}
The first term on the right-hand-side of Eq.~(\ref{barcdmtwo})
dominates in the small-scale limit. Negative $\alpha$ causes
instabilities of density perturbations. Conversely, there are
oscillations of perturbations for positive $\alpha$. In order to
avoid the presence of negative sound speed, we will require
$\alpha\geq0$ in this case.

We find that the perturbations for the barotropic model,
Eqs.~(\ref{cenergy-sg})-(\ref{barcdmtwo}), obey
the same dynamical evolutions as those for the unified GCG model,
Eqs.~(\ref{deltagcg}), (\ref{thetagcg}) and (\ref{gcgtwodot}), where
we identify
\begin{eqnarray}
\label{gCg-dm}
\delta_{gCg}=\frac{\rho_{dm}}{\rho_{dm}+(1+\alpha)V}\delta_{dm} \,,
~~~~~\theta_{gCg}=\theta_{dm} \,.
\end{eqnarray}
We should not be surprised that this interacting model reproduces
the same behaviour as the GCG since by requiring $V=V(\rho_{dm})$ we
have set the pressure of the interacting matter+vacuum $P_{dm+V}=-V$
to be an implicit function of $\rho_{dm}+V$, i.e., we have a
barotropic fluid.

\begin{figure}[!htbp]
\includegraphics[width=12cm,height=6cm]{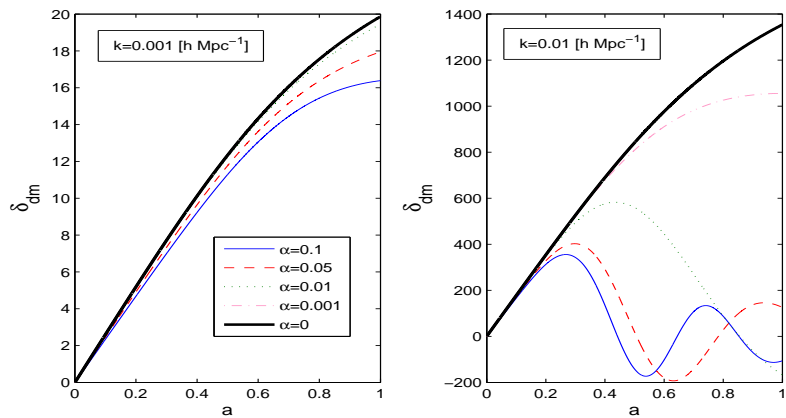}\\
  \caption{The dark matter density perturbations on fixed scales: k = 0.001 $[\textmd{h~Mpc}^{-1}]$ (left panel) and
  k = 0.01 $[\textmd{h~Mpc}^{-1}]$ (right panel) as a function of scale factor for the barotropic model. The thin solid line, dashed line, dotted line, dot-dashed line and thick solid line correspond to $\alpha$ = 0.1, 0.05, 0.01, 0.001, 0 ($\Lambda$CDM model), respectively.
 }\label{fig:addeltac}
\end{figure}

The evolution of the dark matter density perturbations for the
barotropic model on a fixed scale is shown in
Fig.~\ref{fig:addeltac}. In the left panel, it is shown that
increasing $\alpha$ causes the suppression of the growth of density
fluctuations on the scale, $k=0.001[\textmd{h~Mpc}^{-1}]$. For the
smaller scale, $k=0.01[\textmd{h~Mpc}^{-1}]$, in the right panel, we
can see that density fluctuations rapidly decay and oscillate with
increasing $\alpha$.

\subsubsection{Geodesic model}

An alternative model for the energy transfer is to consider an energy flow that is parallel to the four-velocity of the dark matter:
\begin{equation}
Q^{\mu}_{(dm)} = -Qu^{\mu}_{(dm)}.
\end{equation}
For this form of the covariant interaction there is no momentum
transfer in the rest frame of dark matter \cite{Valiviita:2008,
Koyama:2009}, $f_{dm}-Q(\theta-\theta_{dm})=0$ in Eq.~(\ref{QmuI}).
The velocity perturbation for dark matter is not affected by the
interaction and obeys the standard equation, $\dot{\theta}_{dm}=0$
from Eq.~(\ref{fdm}). This allows us to work in a synchronous gauge
that is comoving with the dark matter ($\theta_{dm}=0$). The dark
matter four-velocity, $u^\mu$ is thus a geodesic flow:
$u_{(dm)}^\mu\nabla_\mu u_{(dm)}^\nu=0$. Hence we will refer to this
form of energy-transfer as a geodesic model.

The vacuum perturbation in the dark matter-comoving frame is
identically zero, from Eq.~(\ref{finatheta}),
\begin{eqnarray}
\delta V_{com} = \delta V+\dot{V}\theta_{dm} =0 \,.
\end{eqnarray}
The pressure perturbation for the interacting vacuum+dark matter is
therefore zero in the dark matter rest frame and thus the speed of
sound for the interacting vacuum+dark matter is zero\footnote{A
vanishing pressure perturbation was also considered in a Cardassian
model without any internal fluctuations, which in a special case
that $\nu=1$ reproduces a GCG background \cite{Koivisto:2005}.
However we will show that in the geodesic model dark matter density
perturbations are affected by the background interaction and obey a
modified perturbation equation.} \cite{Wands:2012, inprep}. Note
that this implies that in this case there is a non-adiabatic
pressure perturbation, due to the relative entropy perturbation
between the dark matter and the vacuum energy \cite{Wands:2012,
inprep}, in contrast to the barotropic model.

From Eq.~(\ref{finadelta}), the perturbation equation for the dark
matter density contrast in the rest frame of dark matter obeys
\begin{eqnarray}
 \dot{\delta}_{dm}=-\frac{\dot{h}}{2}+\frac{Q}{\rho_{dm}}\delta_{dm} \,.
\end{eqnarray}
We see that, again, the case with $\alpha=0$ is completely equivalent to $\Lambda$CDM at linear order.

A second-order differential equation for the dark matter overdensity
can be derived from the above equation and Eq.~(\ref{htwodot}),
which gives
\begin{eqnarray}
\label{geocdmtwo}
\ddot{\delta}_{dm}+\left(-\frac{Q}{\rho_{dm}}+2H\right)\dot{\delta}_{dm}-\left[
2H\frac{Q}{\rho_{dm}}+\dot{\left(\frac{Q}{\rho_{dm}}\right)}\right]\delta_{dm}=4
\pi G (\delta\rho+3\delta P) \,.
\end{eqnarray}

\begin{figure}[!htbp]
\includegraphics[width=12cm,height=6cm]{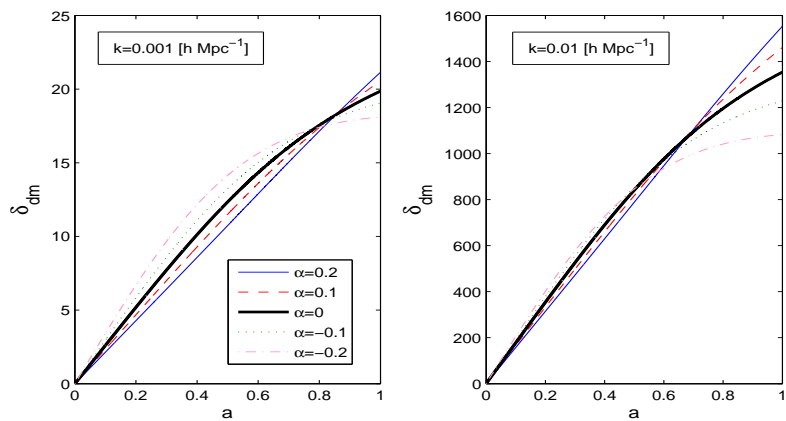}\\
  \caption{The dark matter density perturbations on fixed scales: k = 0.001 $[\textmd{h~Mpc}^{-1}]$ (left panel) and
  k = 0.01 $[\textmd{h~Mpc}^{-1}]$ (right panel) as a function of scale factor for the geodesic model. The thin solid line, dashed line, thick solid line, dotted line and dot-dashed line correspond to $\alpha$ = 0.2, 0.1, 0 ($\Lambda$CDM model), -0.1, -0.2, respectively.
 }\label{fig:naddeltac}
\end{figure}

We plot the evolution of the dark matter density perturbations on
different scales in Fig.~\ref{fig:naddeltac}. Compared with the
growth in the $\Lambda$CDM model, positive $\alpha$ indicates that
there is an energy transfer from dark matter to vacuum energy, so
the growth of density perturbations with positive $\alpha$ is
suppressed at early times during matter-domination. At late times,
vacuum energy dominates in the Universe, and the vacuum interaction
drives the growth of dark matter density perturbations. Conversely,
for the negative $\alpha$ the growth of density perturbations is
enhanced at early times and is gradually suppressed at late times.
Note that there is no instability associated with imaginary speed of
sound for $\alpha<0$ in the geodesic model.

\section{Cosmological constraints}

\subsection{Theoretical predictions of CMB and LSS power spectra}

First, we illustrate the imprint on the CMB and LSS power spectra
for different values of the parameter $\alpha$ for both the
barotropic model and geodesic model. For the barotropic model we
will restrict the value of $\alpha$ to be greater than or equal to
zero in order to avoid instabilities due to an imaginary sound
speed, but we will extend $\alpha$ to negative values for the
geodesic model since the sound speed remains real in that case.

\begin{figure}[!htbp]
\includegraphics[width=8.3cm]{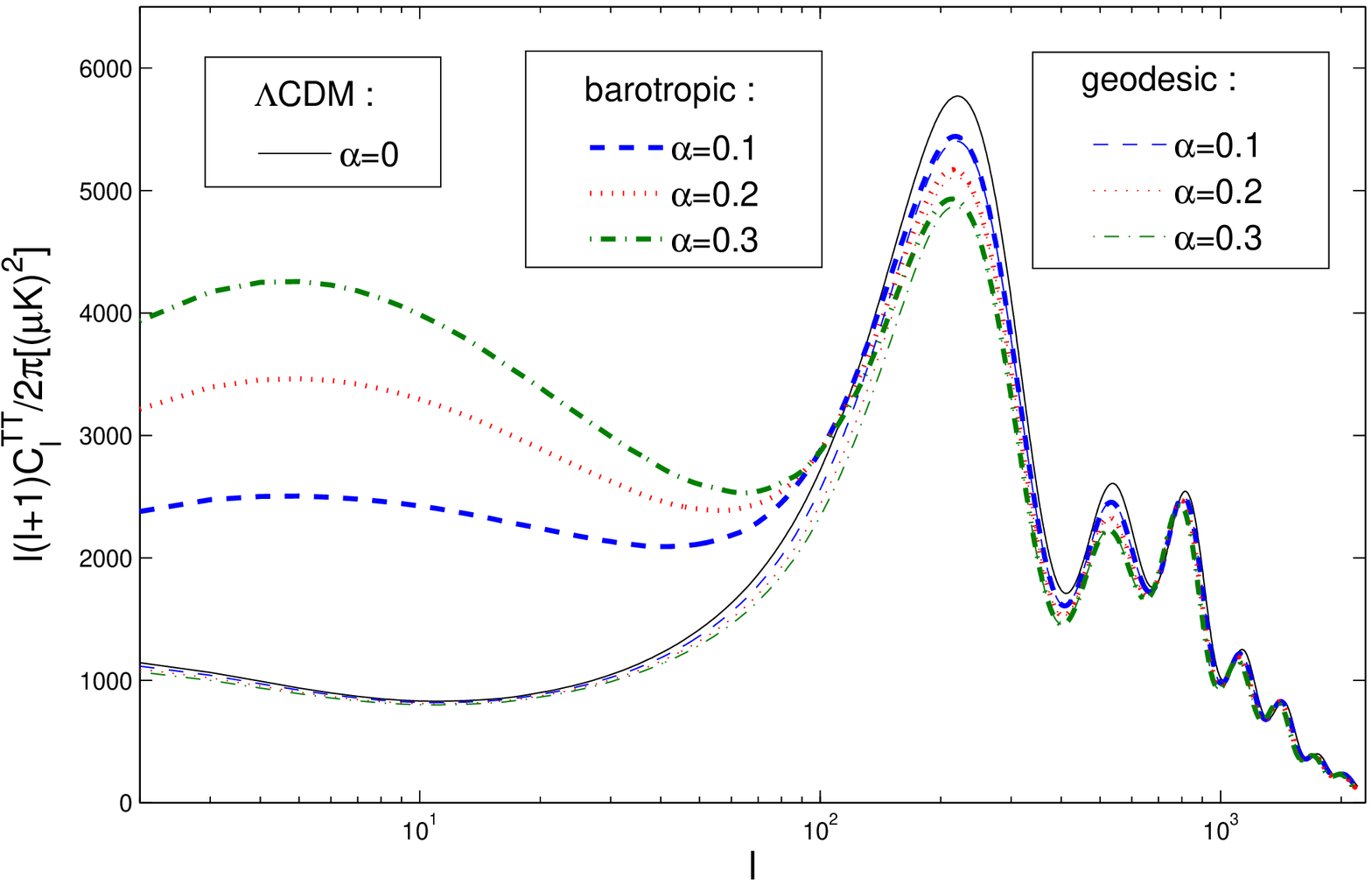}
\includegraphics[width=8.3cm]{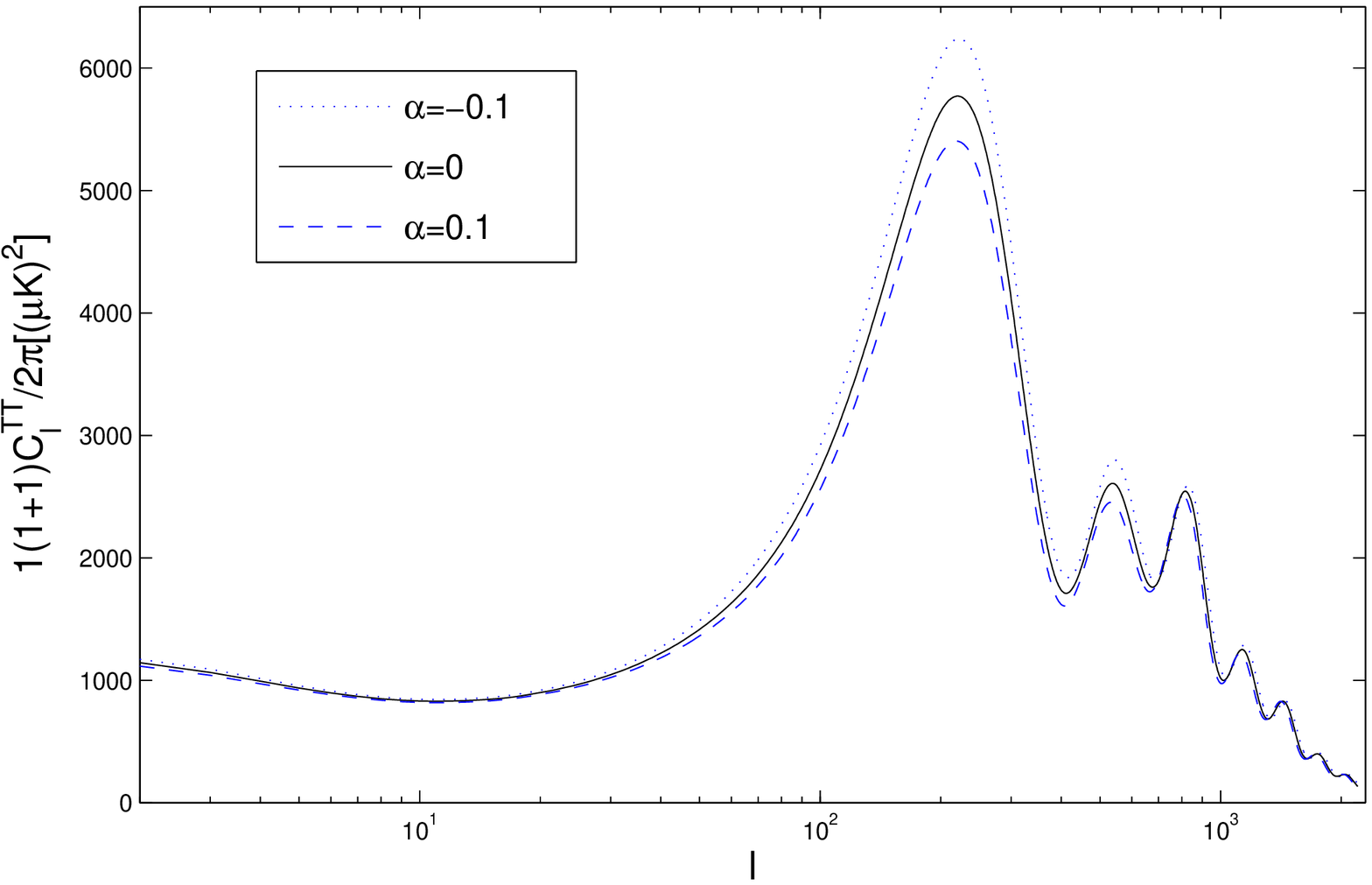}\\
  \caption{CMB temperature power spectra vs multipole moment l.
The solid line is for the $\Lambda$CDM model. Left panel:
The thick lines and thin lines for the same values of parameter
$\alpha$ correspond to the barotropic model and geodesic model,
respectively. Right panel: The dashed line and dotted line
correspond to $\alpha = 0.1$ and $\alpha = -0.1$ in the geodesic
model, respectively. }\label{fig:cls}
\end{figure}

The effects on the CMB power spectra are shown in
Fig.~\ref{fig:cls}. From the left panel, positive $\alpha$ leads to
an energy transfer from dark matter to vacuum energy. Increasing
$\alpha$ causes the matter-radiation equality to happen earlier and
hence suppresses the first acoustic peaks. Conversely, negative
$\alpha$ delays the epoch of matter-radiation equality, and hence
amplifies the acoustic peaks, as shown in the geodesic case in the
right panel of Fig.~\ref{fig:cls}.

One source of CMB anisotropies at low multipoles is the ISW
effect. The amplitude of the ISW signal on large scales is affected not only
by the epoch of dark matter-vacuum energy equality,
but is also modified directly by the evolution of the gravitational potential, which differs
in the barotropic and geodesic models. These two competing mechanisms give
the final ISW signal.

In order to clearly express the ISW effect caused by the
time-varying gravitational potential, we calculate the Newtonian
potential, $\psi^N$, corresponding to the metric perturbation,
$\psi$, in the Newtonian or longitudinal gauge ($B=E=0$)
\cite{ref:Ma1995}.

This is given by the relativistic Poisson equation
\cite{Malik:Wands:review}
\begin{eqnarray}
 \label{poisson}
k^2\psi^N=-4\pi G a^2 \rho\Delta \,,
\end{eqnarray}
where the total comoving density perturbation is given by
\begin{eqnarray}
\rho\Delta=\sum_I\rho_I[\delta_I-3H(1+w_I)\theta_I] \,,
\end{eqnarray}
including the vacuum perturbation.
The traceless, space-space component of the Einstein field equations,
(\ref{traceless}), gives the relation between the two scalar
potentials in the Newtonian gauge \cite{Malik:Wands:review}:
\begin{eqnarray}
 \label{potential-relation}
\psi^N-\phi^N=8 \pi G \Pi \,,
\end{eqnarray}
where $\Pi$ is related to total the anisotropic stresses.

\begin{figure}[!htbp]
\includegraphics[width=12cm]{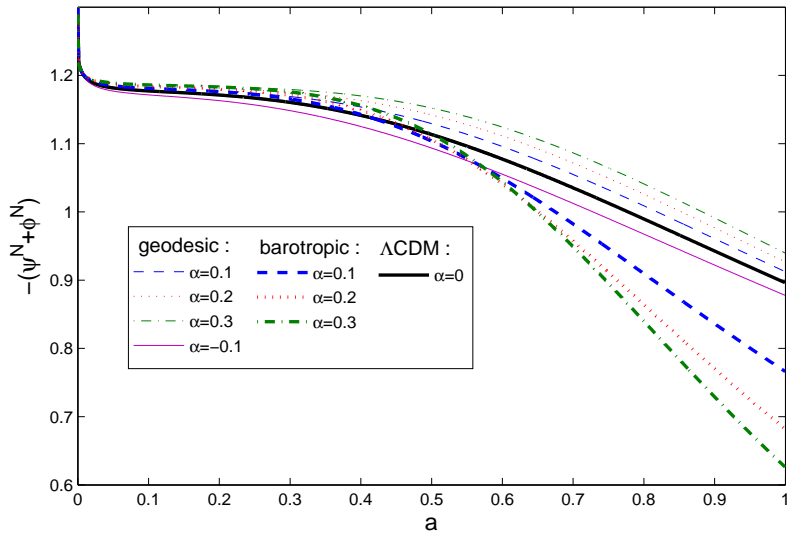}\\
  \caption{The evolution of the gravitational potential as a
function of scale factor on the large scale $k=0.001 [\textmd{h
Mpc}^{-1}]$. The thick solid line is for the $\Lambda$CDM model.
The thick dashed, dotted and dot-dashed lines correspond to $\alpha$ = 0.1, 0.2, 0.3 for the barotropic model,
respectively. The thin dashed, dotted and dot-dashed lines correspond to $\alpha$ = 0.1,
0.2, 0.3 for the geodesic model, respectively. The thin solid line is the evolution of the
gravitational potential for $\alpha = -0.1$ in the geodesic model.
%
%
 }\label{fig:source}
\end{figure}

Now the ISW temperature anisotropy is given by the following source:
\begin{eqnarray}
 \label{iswsource}
S_{ISW}&=& \dot{\psi}^N+\dot{\phi}^N \nonumber \\
&=& -8\pi G \frac{d}{dt} \left ( \frac{a^2}{k^2} \rho\Delta + \Pi \right) \,.
\end{eqnarray}
At late times, the relativistic particles (photons and neutrinos)
can be neglected and there is no anisotropic stress. Therefore, it
is the comoving density perturbation that drives the late-time ISW
effect. Figure \ref{fig:source} shows how the parameter $\alpha$ affects
the time evolution of the gravitational potential in the two models.
We see that for the same value of $\alpha$ we find a much
larger variation in the Newtonian potential at late times in the
barotropic model due to the nonzero sound speed, as shown in
Fig.~\ref{fig:cs}. This will be reflected in much tighter bounds on
the value of $\alpha$ in the barotropic model coming from the
overall CMB anisotropies, and these constraints become even
tighter when the galaxy clustering data is also used to identify the
late-time ISW effect.

Note that the overall CMB anisotropies, including the ISW effect,
are the same in the barotropic model and the original unified
Chaplygin gas model\footnote{This case can also be obtained from
case II of the Cardassian model in Ref.~\cite{Koivisto:2005} for
$\nu=1$.} since the evolution of the total matter+vacuum density
perturbations is the same as the density perturbations in the
unified GCG model.

\begin{figure}[!htbp]
\includegraphics[width=16cm]{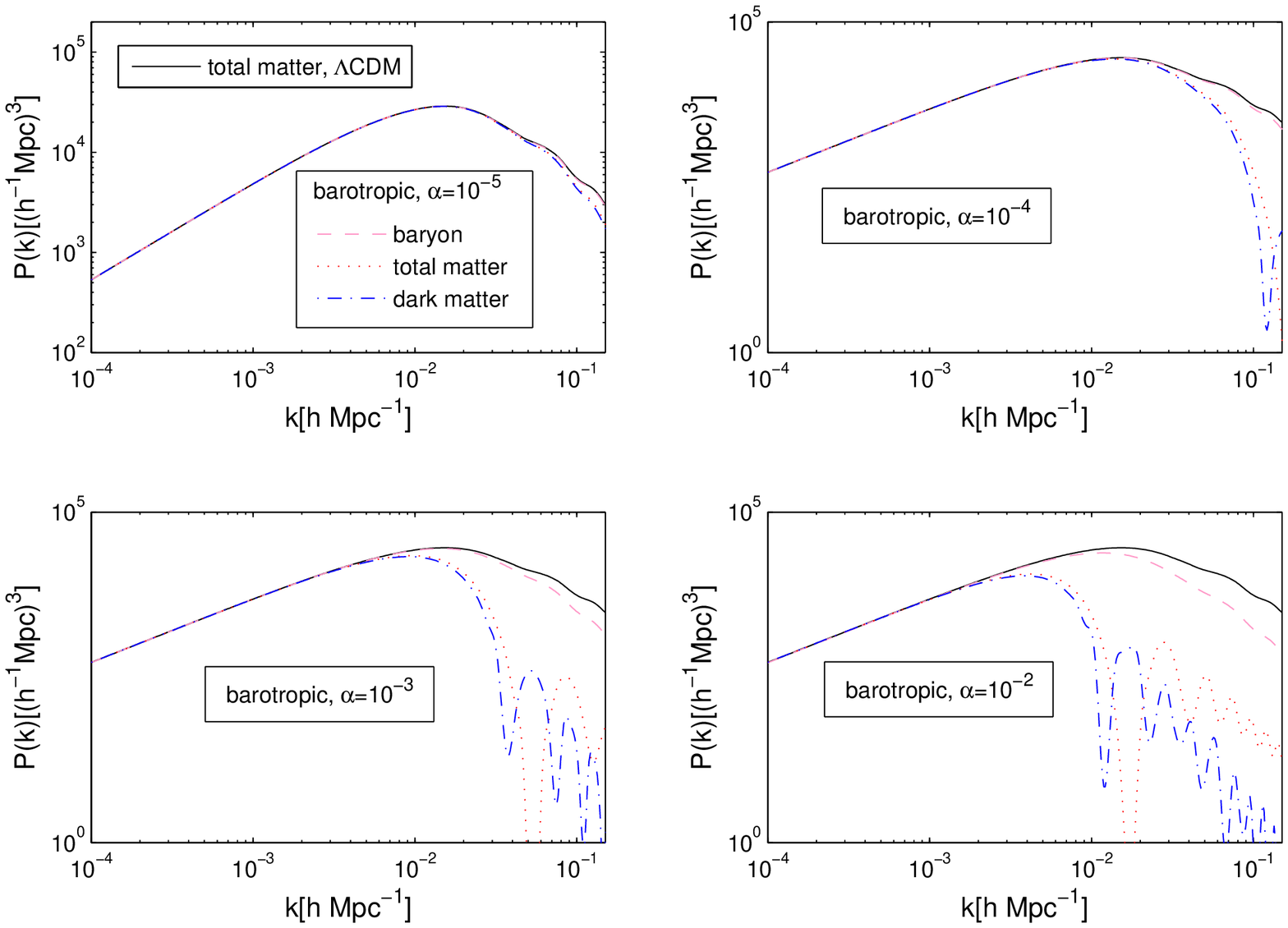}\\
  \caption{The LSS power spectra as a function of wavenumber for the
barotropic model. In each panel, the solid line is the total matter
power spectrum for the $\Lambda$CDM model. The dashed line, dotted line
and dot-dashed line are the baryonic power spectrum, the total matter
power spectrum and the dark matter power spectrum, respectively. The
separate panels (left to right and top to bottom) correspond to
$\alpha = 10^{-5}, 10^{-4}, 10^{-3}, 10^{-2}$, respectively.
 }\label{fig:adPk}
\end{figure}

In Fig.~\ref{fig:adPk}, we present the dark matter power spectrum,
the baryonic power spectrum, and the total matter power spectrum
(baryon and dark matter) for the barotropic model. The figure shows
that, as expected, the dark matter
has an oscillating power spectrum in the small-scale limit for
nonzero $\alpha$. The larger the parameter $\alpha$, the larger the
scale where the oscillations of dark matter power spectra begin. By
comparison, the baryonic power spectra behave much more smoothly as
they are only gravitationally coupled to the oscillating dark matter
density \cite{Park:2009np}. The total matter power spectra combining
both dark matter and baryon densities therefore shows intermediate
behaviour, but cannot avoid the rapid decay of power spectrum on the
small scales for nonzero $\alpha$.

\begin{figure}[!htbp]
\includegraphics[width=12cm]{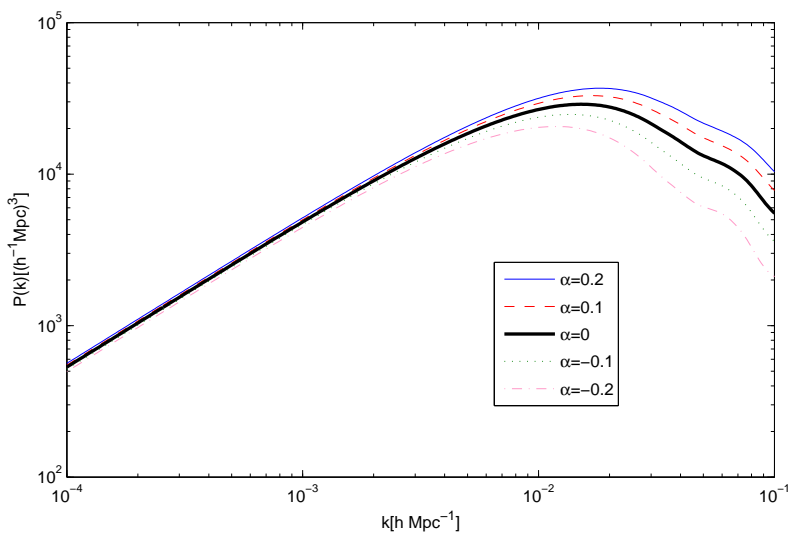}\\
  \caption{The total matter power spectra as a function of wavenumber for the
geodesic model. The thin solid line, dashed line, thick solid line, dotted line and
dot-dashed line correspond to $\alpha$ = 0.2, 0.1, 0 ($\Lambda$CDM model),
-0.1, -0.2, respectively.
 }\label{fig:nadiPk}
\end{figure}

Figure \ref{fig:nadiPk} shows the total matter power spectra for the
geodesic model. The problematic oscillations of the dark matter
power spectra do not occur in this case since the sound speed is
zero. The only oscillations seen are the baryon acoustic
oscillations, also seen in $\Lambda$CDM. The matter power spectrum
is enhanced on small scales for $\alpha>0$ due to the earlier
matter-radiation equality which moves the turnover in the matter
power spectrum to smaller scales.

\subsection{Constraint method and data}

Next, we will use a variety of cosmological data to constrain these
two models for the matter+vacuum interaction, according to the
Markov Chain Monte Carlo method, using the publicly available
CosmoMC package \cite{ref:MCMC,ref:MCMCweb}. The cosmological
parameter set is $P\equiv\{\Omega_{b}h^2, \Omega_{dm}h^2, \Theta_S,
\tau, \alpha, n_s, \log[10^{10}A_s]\}$, where $\Omega_bh^2$ and
$\Omega_{dm}h^2$ are the physical baryon and dark matter densities,
$\Theta_S$ is the ratio (multiplied by 100) of the sound horizon and
angular diameter distance, $\tau$ is the optical depth, $\alpha$ is
the dimensionless generalised Chaplygin gas model parameter, $n_s$
is the scalar spectral index, and $A_s$ is defined as the amplitude
of the initial power spectrum. Note that the dimensional parameter
$A$ in the original definition of the Chaplygin gas (\ref{PgCg}) is
given in terms of $\Omega_{dm}$ and $\Omega_V$ using Eq.~(\ref{A})
as
\begin{equation}
 A = \Omega_V \left( \Omega_{dm} + \Omega_V \right)^\alpha \left( \frac{3H_0^2}{8\pi G} \right)^{1+\alpha} \,,
\end{equation}
and $\Omega_V=1-\Omega_b-\Omega_r-\Omega_{dm}$, by
the requirement of spatial flatness.

We will use different combinations of data (CMB+SNIa+BAO, CMB+SNIa+LSS, and
CMB+SNIa+gISW). For the LSS power spectra we will only
use the linear matter power spectrum up to $k\sim0.1 [h
~\textmd{Mpc}^{-1}]$.
In addition, we use a top-hat prior for the cosmic age, i.e., $10
\textmd{Gyr}<t_0<20 \textmd{Gyr}$, impose a weak Gaussian prior on
the physical baryon density $\Omega_{b}h^2=0.022\pm0.002$ from big
bang nucleosynthesis \cite{ref:bbn}, and use the value of the Hubble
constant $H_0=73.8\pm2.4 {\rm km~s}^{-1} {\rm Mpc}^{-1}$
\cite{ref:H0} taking a Gaussian likelihood function.

The theoretical galaxy-ISW (gISW) cross-correlated power spectrum,
$P_{\delta,\dot{\Phi}}$, where $\dot{\Phi}\equiv
\dot{\psi}^N+\dot{\phi}^N$, can be calculated by a modified CAMB code. We
modify the code by referring to ISiTGR \cite{ref:ISiTGR}, so that a
power spectrum can be directly calculated from the individual
transfer function. In the absence of magnification bias, the gISW
cross correlation is written as
\begin{eqnarray}
C_l^{g_xT}=\frac{T_{CMB}}{(l+1/2)^2}\int dz
b_x(z)\Pi_x(z)P_{\delta,\dot{\Phi}}\left(\frac{l+1/2}{\chi(z)},z\right)
\,,
\end{eqnarray}
where the galaxy samples in Ref.~\cite{ref:gISW} are collected from the
Two Micron All Sky Survey (2MASS) Extended Source Catalog (XSC),
which is divided into 4 bins: 2MASS0-3, the luminous red galaxies
(LRGs) from SDSS split into 2 sample bins: LRG0-1, the photometric
quasars from SDSS with 2 redshift bins: QSO0-1, and NRAO VLA Sky
Survey (NVSS): NVSS. Namely, there are nine galaxy sample bins $x$.
The scale-independent bias factor $b_x(z)$ is assumed to relate the
observed projected galaxy overdensity to the matter density.
$\Pi_x(z)$ is the redshift distribution. $\chi(z)$ is the comoving
radial distance. In ISWWLL code, the function $f_x(z)\equiv
b_x(z)\Pi_x(z)$ is determined by fitting auto power spectra and
cross power spectra between the samples:
\begin{eqnarray}
C_l^{g_xg_y}=\int dz
f_x(z)f_y(z)\frac{H(z)}{\chi^2(z)}D^2(z)P_{\delta,\delta}\left(\frac{l+1/2}{\chi(z)},0\right)
\,,
\end{eqnarray}
even if the methods differ for the various samples.

For the 2MASS galaxies, the redshifts are identified by matching
2MASS galaxies with SDSS main galaxies. The nonlinear $\textbf{Q}$
model \cite{Qmodel} (denoted by the bold text to be distinguished
from the background interaction $Q$) of the galaxy power spectrum is
used to get the bias $b(z)$ and parameter $\textbf{Q}$ by fitting
the auto power spectra of the galaxies to the measured values. For
the SDSS LRGs, the redshift distribution is determined by the method
in Ref.~\cite{redshiftdis}, and the bias is also obtained by adopting the
same $\textbf{Q}$ model fitting to the measurements as done for the
2MASS samples. For the SDSS quasars, the preliminary redshift
distribution is constructed and is used to predict the shape of the
function $f_{QSO0-1}(z)$ in the presence of magnification bias. For
NVSS, the analytic $\Gamma$ distribution for $f_{NVSS}(z)$ is
constrained by NVSS auto power spectrum and cross power spectra with
other samples.

For each sample, the individual $f_x(z)$ need to be recomputed when
the cosmological parameters are changed. We have modified the code
\cite{ref:gISW} to give the correct background evolution for our
case, following Eq.~(\ref{finalFriedmann}), and the second-order
differential equation for the total matter density perturbation is
also modified due to the interaction of the dark matter density
perturbation with the vacuum, as given in Eqs.~(\ref{barcdmtwo})
or~(\ref{geocdmtwo}) for our two interaction models. The total
matter density contrast is given by
\begin{eqnarray}
\delta_m(z)=\frac{\rho_{dm}\delta_{dm}+\rho_b\delta_b}{\rho_{dm}+\rho_b} \,.
\end{eqnarray}
We have checked that compared with the density perturbation, the
dark matter velocity term, $\theta_{dm}$, on the right-hand-side of
Eq.~(\ref{barcdmtwo}) is small enough to be ignored when we solve
the second-order differential equation (\ref{barcdmtwo}) for the
barotropic case.

\subsection{Barotropic model constraints}

\begin{figure}[!htbp]
\includegraphics[width=16cm]{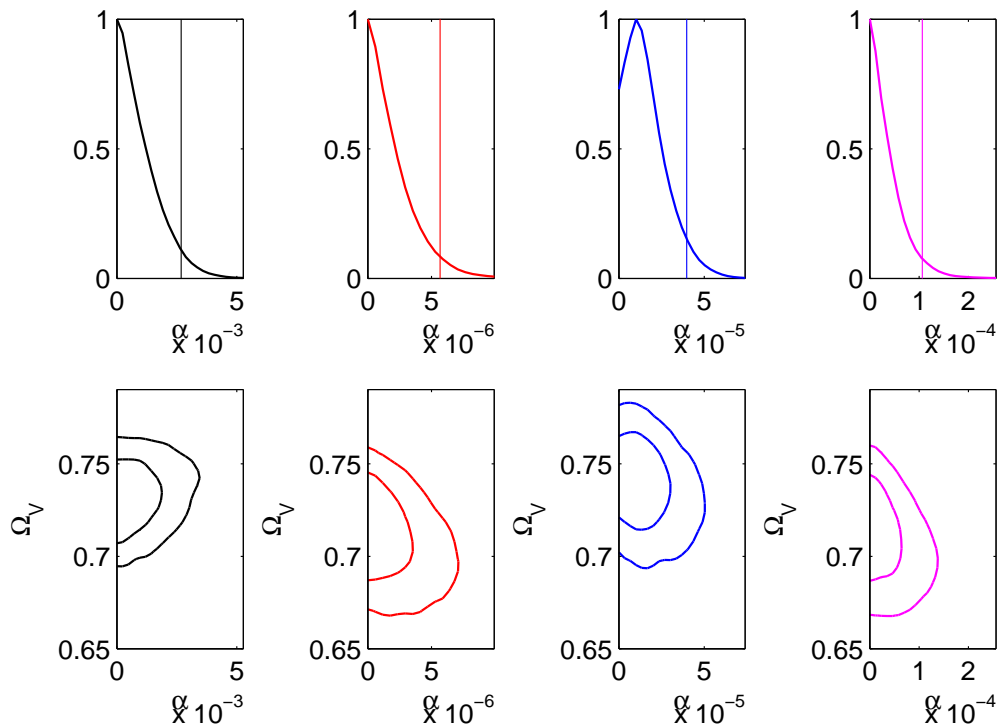}\\
  \caption{ The 1D marginalized distributions for the parameter $\alpha$ and 2D contours with 68\% C. L. and 95\% C. L.
for the barotropic model using the combinations of data (left to right): CMB+BAO+SNIa
(black line), CMB+SNIa+LSS(m) (red line), CMB+SNIa+gISW (blue line)
and CMB+SNIa+LSS(b) (magenta line), respectively. The vertical lines
in the top panels are the 95\% confident levels.
 }\label{fig:ad1d2d}
\end{figure}

We present the constraints for the barotropic model from the
combined data sets of CMB+SNIa+BAO, CMB+SNIa+LSS(m), CMB+SNIa+gISW,
and CMB+SNIa+LSS(b) in Table \ref{tab:barotropic-results}. Here
LSS(b) denotes the constraints obtained by fitting the LSS power spectra to the baryonic matter
power spectra, while LSS(m) means that the theoretical
predictions for the total matter (baryon+dark matter) power spectra
are used. The corresponding 1D marginal distribution of the
parameter $\alpha$ and the 2D contour plots of the parameters
$\alpha$ and $\Omega_V$ are shown in Fig.~\ref{fig:ad1d2d}.

From Table \ref{tab:barotropic-results} and
Fig.~\ref{fig:ad1d2d}, we can see that the joint data from
CMB+SNIa+BAO can constrain $\alpha$ to be less than or of order
$10^{-3}$. Tighter bounds on the parameter $\alpha$ can be obtained
by using the LSS power spectra data. When the total matter power
spectra are used to fit the SDSS DR7 power spectra data combined
with data from CMB+SNIa, the tightest bounds on the parameter
$\alpha$ are obtained, to less than or of order $10^{-6}$. By
comparison when the baryonic power spectra are are fitted to the
SDSS DR7 power spectra data, larger values of $\alpha$ are allowed
and the upper limit at 95\% C. L. is
$1.07\times10^{-4}$. This is because the dark matter power spectra
are damped on small scales more quickly than the baryonic power
spectra
as shown in Fig.~\ref{fig:adPk}. From the thick lines in the left
panel of Fig.~\ref{fig:cls}, it is seen that the low-multipole CMB
power spectrum for the barotropic model is obviously enhanced with
the increasing $\alpha$ through the ISW effect. By using the data of
CMB+SNIa+gISW, we obtain $\alpha<3.97\times10^{-5}$ at the 95\% C.
L.

\begin{table}
\begin{center}
\begin{tabular}{|cc|   cc|   cc|  cc|  cc|}
\hline \hline Parameters &&CMB+SNIa+BAO& & CMB+SNIa+LSS(m)& &
CMB+SNIa+gISW& & CMB+SNIa+LSS(b)&
\\ \hline
$\Omega_b h^2$ &&  $        0.0225_{-        0.0005}^{+ 0.0005}$ &
 &$        0.0224_{-        0.0005}^{+        0.0005}$   &
 &$        0.0224_{-        0.0005}^{+        0.0005}$  &
 &$        0.0224_{-        0.0005}^{+        0.0005}$  &\\
$\Omega_{dm} h^2$ &&  $        0.1115_{-        0.0032}^{+ 0.0033}$
 & &$        0.1156_{-        0.0035}^{+        0.0035}$  &
&$        0.1102_{-        0.0038}^{+        0.0038}$   &
&$        0.1156_{-        0.0034}^{+        0.0035}$ &\\
$\Theta_S$ && $        1.0396_{-        0.0025}^{+        0.0025}$
 & & $        1.0396_{-        0.0025}^{+        0.0025}$  &
& $        1.0395_{-        0.0025}^{+        0.0025}$   &
&$        1.0395_{-        0.0025}^{+        0.0025}$  &\\
$\tau$ && $        0.0872_{-        0.0067}^{+        0.0060}$
  & &$        0.0851_{-        0.0072}^{+        0.0062}$&
&$        0.0880_{-        0.0070}^{+        0.0064}$  &
& $        0.0855_{-        0.0066}^{+        0.0061}$ &\\
$\alpha$ && $   ~<2.68\times10^{-3}$ & & $ ~<5.66\times10^{-6}$ &
&$~<3.97\times10^{-5}$&
&$~<1.07\times10^{-4}$&\\
$n_s$ &&  $        0.970_{-        0.012}^{+        0.013}$ & &  $
0.964_{-        0.012}^{+        0.012}$ & &$        0.967_{-
0.012}^{+        0.012}$  && $        0.964_{-        0.012}^{+        0.012}$  &\\
$\textmd{log}[10^{10} A_s]$ && $        3.081_{-        0.033}^{+
0.033}$ & & $        3.090_{-        0.032}^{+        0.032}$ & &$
3.075_{-        0.034}^{+        0.034}$  &
&$        3.091_{-        0.032}^{+        0.032}$ &\\
$\Omega_V$ && $        0.73_{-        0.01}^{+        0.01}$
 & &$        0.71_{-        0.02}^{+        0.02}$
 & &$        0.74_{-        0.02}^{+        0.02}$
 & &$        0.71_{-        0.02}^{+        0.02}$  &\\
$H_0$ && $       70.8_{-        1.2}^{+        1.3}$
 & &  $       69.1_{-        1.5}^{+        1.5}$
 & & $       71.2_{-        1.6}^{+        1.5}$
 & & $       69.1_{-        1.5}^{+        1.5}$  &\\
\hline \hline
\end{tabular}
\caption{The mean values with $1\sigma$ errors and marginalized 95\%
C. L. of parameter $\alpha$ for the barotropic model using different
combinations of data sets.
}\label{tab:barotropic-results}
\end{center}
\end{table}

\subsection{Geodesic model constraints}

\begin{figure}[!htbp]
\includegraphics[width=12cm,height=8cm]{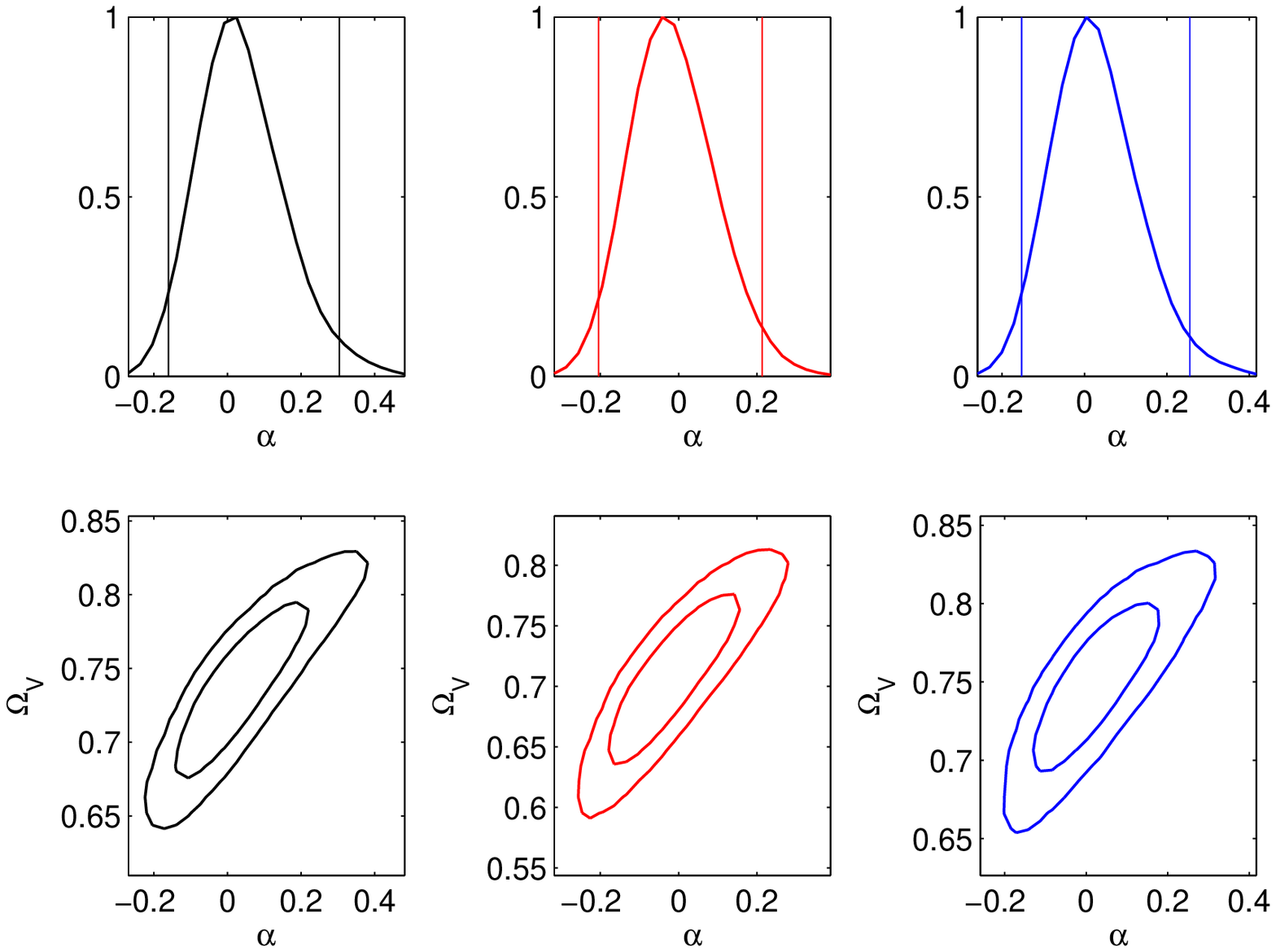}\\
  \caption{ The 1D marginalized distributions for the parameter $\alpha$ and 2D contours with 68\% C. L. and 95\% C. L.
for the geodesic model using the combinations of data (left to right): CMB+SNIa+BAO (black
line), CMB+SNIa+LSS(m) (red line), and CMB+SNIa+gISW (blue line),
respectively. The vertical lines in the top panels are the 95\%
confident levels. }\label{fig:geo1d2d}
\end{figure}

For the geodesic model, the constraints from the combination of
various data are listed in Table \ref{tab:geodesic-results}. The
marginal 1D distributions and 2D contour plots from CMB+SNIa+BAO,
CMB+SNIa+LSS(m), and CMB+SNIa+gISW are shown in
Fig.~\ref{fig:geo1d2d}.

CMB+SNIa+LSS(m) constraints allow negative mean value for $\alpha$.
Compared with results from CMB+SNIa+BAO and CMB+SNIa+LSS(m)
constraints, CMB+SNIa+gISW gives a relatively narrow allowed range
for $\alpha$. The allowed range of the parameter $\alpha$ at the
95\% C. L. in this case are $-0.16<\alpha<0.30$ from CMB+SNIa+BAO,
$-0.20<\alpha<0.21$ from CMB+SNIa+LSS(m) and $-0.15<\alpha<0.26$
from CMB+SNIa+gISW, respectively. By comparison with the results for
the barotropic model on the basis of the same data sets, the allowed
region of parameter $\alpha$ for the geodesic model is obviously
enlarged.

\begin{figure}[!htbp]
\includegraphics[width=18cm]{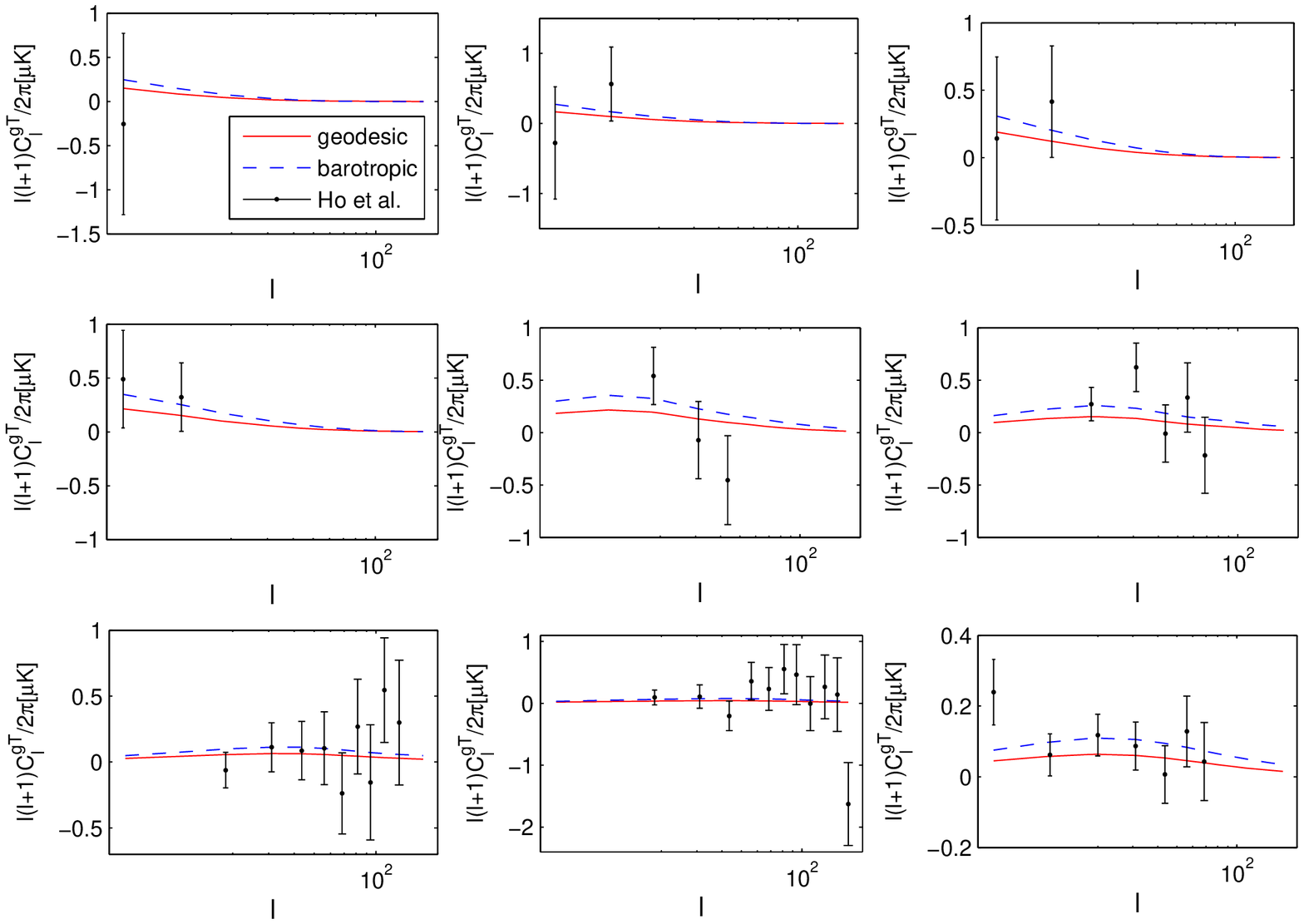}\\
  \caption{ The gISW cross-correlation power spectra with different galaxy
samples for the barotropic model (dashed line) with mean values taken
from CMB+SNIa+gISW constraints in the Table
\ref{tab:barotropic-results} and the geodesic model (solid line) with
mean values taken from CMB+SNIa+gISW constraints in the Table
\ref{tab:geodesic-results}, respectively. The dots with error
bars denote the gISW data from Ho et al. \cite{ref:gISW}.
}\label{fig:clgisw}
\end{figure}

Finally, in Fig.~\ref{fig:clgisw} we show gISW cross-correlation
power spectra predicted by using the mean values of CMB+SNIa+gISW
for the two models.
\begin{table}
\begin{center}
\begin{tabular}{|cc|   cc|   cc|  cc|}
\hline \hline Parameters &&CMB+SNIa+BAO& & CMB+SNIa+LSS(m)& &
CMB+SNIa+gISW&
\\ \hline
$\Omega_b h^2$ && $    0.0224_{-    0.0005}^{+ 0.0005}$& &  $
0.0224_{-        0.0005}^{+        0.0005}$    &
& $    0.0225_{-    0.0005}^{+    0.0005}$ & \\
$\Omega_{dm} h^2$ & &$    0.1077_{-    0.0146}^{+ 0.0144}$& & $
0.1167_{-        0.0150}^{+        0.0149}$   &
&$        0.1062_{-        0.0135}^{+        0.0135}$ &\\
$\Theta_S$ && $    1.0395_{-    0.0026}^{+ 0.0026}$ & & $
1.0394_{-        0.0026}^{+        0.0026}$   &
&$        1.0397_{-        0.0025}^{+        0.0025}$ &\\
$\tau$ && $    0.0864_{-    0.0072}^{+    0.0063}$& &$
0.0862_{-        0.0072}^{+        0.0063}$  &
& $        0.0895_{-        0.0076}^{+        0.0065}$  & \\
$\alpha$ && $    0.04_{-    0.11}^{+    0.11}$& &  $ -0.01_{-
0.11}^{+        0.11}$   &
&$        0.03_{-        0.10}^{+        0.10}$  &\\
$n_s$ && $    0.964_{-    0.013}^{+    0.013}$& & $ 0.965_{-
0.013}^{+        0.013}$   &
& $        0.967_{-        0.013}^{+        0.013}$  & \\
$\textmd{log}[10^{10} A_s]$ && $        3.081_{-        0.033}^{+
0.033}$ & & $        3.086_{-        0.033}^{+        0.033}$   &
&  $        3.076_{-        0.034}^{+        0.034}$ & \\
$\Omega_V$ && $        0.74_{-        0.04}^{+        0.04}$ & &$
0.71_{-        0.04}^{+        0.04}$  &
& $        0.75_{-        0.03}^{+        0.03}$ & \\
$H_0$ &&  $       70.6_{-        1.5}^{+        1.5}$ & & $ 69.3_{-
1.9}^{+        1.9}$  &
&$       71.8_{-        1.7}^{+        1.7}$  &\\
\hline \hline
\end{tabular}
\caption{The mean values with $1\sigma$ errors for the geodesic
model using different combinations of data
sets.}\label{tab:geodesic-results}
\end{center}
\end{table}

\section{Conclusions}

Any unified dark matter model can be decomposed into dark matter interacting with vacuum energy \cite{Bento:2004uh,Wands:2012,inprep}. In such a decomposed model, the vacuum energy is coupled to dark matter and the energy-momentum
transfer between the vacuum energy and the dark matter is given by the gradient of vacuum energy. Different covariant forms for the energy-momentum transfer four-vector have different effects on the CMB angular power spectrum and LSS power spectrum through the different evolution of inhomogeneous perturbation evolutions, even if the background cosmology is the same.
Observational constraints on these decomposed models correspond to constraints on the interaction between dark matter and the vacuum energy, where we recover $\Lambda$CDM in the limit of zero interaction.

In this paper, we studied the evolution of perturbations in a decomposed Chaplygin gas model and compared this with the evolution in a unified Chaplygin gas model with adiabatic perturbations.
We considered two different models for the energy-momentum transfer in the decomposed case. In one model the energy-momentum transfer is along the gradient of the local dark matter density, in which case we recover the evolution equations for perturbations of a single barotropic fluid with the same equation of state as the generalised Chaplygin gas. The pressure perturbations are adiabatic and the rest-frame sound speed equals the adiabatic sound speed. This nonzero sound speed has an effect on
both the CMB power spectrum and the LSS power spectrum. Increasing $\alpha$ enhances
the CMB power spectra at the low multipoles. In addition, the effect on the total matter (baryon+dark matter) power spectrum caused by the sound speed is that there are decaying
oscillations in the matter power spectrum for positive $\alpha$, as for the generalised Chaplygin gas~\cite{Sandvik:2002jz}.

A different possibility which we considered is that the energy-momentum transfer is along the
dark matter four-velocity, which leads to the dark matter following geodesics. Compared with
the barotropic model, the geodesic model has a relative entropy
perturbation between dark matter and vacuum energy. Thus the sound
speed of the interacting dark matter+vacuum energy in the dark
matter rest frame is zero. There are no oscillations or
instabilities in the total matter power spectrum for positive or
negative $\alpha$.

In the decomposed Chaplygin gas, different interaction models can
lead to very different evolutions of perturbations due to the
different sound speeds. We test whether these two kinds of
decomposed models can be supported or distinguished by current
observations. We have constrained the barotropic model and geodesic
model by using various combinations of data sets, including CMB
constraints, type Ia supernovae, BAO distance measurements,
large-scale structure and the integrated Sachs-Wolfe effect.

For the barotropic model the most stringent constraint on the
$\alpha$ parameter is of the order of $10^{-6}$ from the
combinations of CMB+SNIa+LSS(m), where the theoretical total matter
(baryon+dark matter) power spectrum is fitted to the LSS power
spectrum data. We conclude that a barotropic model must be extremely
close to the $\Lambda$CDM model. In contrast the allowed region for
the $\alpha$ parameter in the geodesic model is much larger, and negative $\alpha$ can also be compatible with the
observations. The geodesic model thus allows significant deviations
from the $\Lambda$CDM model.

In the case of a unified dark matter model it is not clear whether the galaxy power spectrum on large scales should follow the total matter power spectrum, which gives rise to the gravitaitonal potential, or the baryonic matter power spectrum, since only baryons follow geodesics. This choice leads to different predictions for the LSS power spectrum \cite{Park:2009np}. The same ambiguity applies in our barotropic model of interacting vacuum+dark matter, since the dark matter does not follow geodesics in this case, so we present constraints using LSS fitted to both the total matter power spectrum and the baryonic matter power spectrum.
In the geodesic model, however, we only use the total matter power spectrum to fit the LSS power spectrum since both the dark matter and the baryons follow geodesics in this case and we expect the total matter overdensity to be responsible for the formation of collapsed halos.
Nonetheless it would be interesting to study further the process of nonlinear collapse in this model and we hope to return to this question in future work.

\acknowledgements{Y.~W.~would like to thank Timothy Clemson, Lucas
Lombriser, Gong-Bo Zhao for help with the CosmoMC and ISWWLL code.
D.~W.~is supported by STFC Grant No. ST/H002774/1. L.~X.~is supported in part by NSFC under the Grants No. 11275035 and ``the Fundamental Research Funds for the Central Universities" under the Grants No. DUT13LK01. J.~D.-S. is supported by CONACYT Grant 210405. A.~H.~is
supported by World Class University Grant No. R32-2009-000-10130-0
through the National Research Foundation, Ministry of Education,
Science and Technology of Korea. All the numerical computations were
done on the Sciama High Performance Compute cluster, supported by
the ICG SEPNet and University of Portsmouth. Y.~W.~and J.~D.-S.~thank the
ICG, University of Portsmouth for their hospitality.}

\end{document}